\documentclass[12pt]{article}
\usepackage{fancyvrb,epsfig,amssymb,amsmath,bbm,psfrag}

\catcode`\@=11
\DeclareMathAlphabet{\mathpzc}{OT1}{pzc}{m}{it}

\newcommand{\insertfig}[2]{\mbox{\epsfxsize=#1cm \epsfbox{#2.eps}}}

\font\cmss=cmss12 
\def\1{\hbox{{1}\kern-.25em\hbox{l}}}
\def\bfZ{\relax{\hbox{\cmss Z\kern-.4em Z}}}

\def \be  {\begin{equation}}
\def \ee  {\end{equation}}
\def \ba  {\begin{eqnarray}}
\def \ea  {\end{eqnarray}}
\def \baa {\begin{eqnarray*}}
\def \eaa {\end{eqnarray*}}
\def \bb  {\begin {thebibliography} }
\def \eb  {\end{thebibliography}}
\def \lab #1 {\label{#1}}

\def \matrix #1 {\left(\begin{array}{cc} #1 \end{array}\right)}

\newcommand{\as}{\ifmmode\alpha_{\rm s}\else{$\alpha_{\rm s}$}\fi}
\newcommand{\asbar}{\ifmmode\bar{\alpha}_{\rm s}\else{$\bar{\alpha}_{\rm
s}$}\fi}

\newcommand{\ft}[2]{{\textstyle\frac{#1}{#2}}}

\font\cmss=cmss12 
\def\inbar{\,\vrule height1.5ex width.4pt depth0pt}
\def\IC{\relax\hbox{$\inbar\kern-.3em{\rm C}$}}
\def\IZ{\relax{\hbox{\cmss Z\kern-.4em Z}}}
\def\IR{{\hbox{{\rm I}\kern-.2em\hbox{\rm R}}}}

\def\IP{{\hbox{{\rm I}\kern-.2em\hbox{\rm P}}}}
\def\II{\hbox{{1}\kern-.25em\hbox{l}}}

\newbox\lett\newdimen\lheight\newdimen\lwidth
\def\ontop#1#2{\setbox\lett=\hbox{#2}\lheight\ht\lett
\multiply\lheight by 12 \divide\lheight by 10\relax%
\lwidth\wd\lett \multiply\lwidth by 8 \divide\lwidth by 10\relax #2\kern-
\lwidth%
\raise\lheight\hbox{{$\scriptstyle #1$}}\kern.1ex}

\def\XXint#1#2#3{{\setbox0=\hbox{$#1{#2#3}{\int}$}
     \vcenter{\hbox{$#2#3$}}\kern-.5\wd0}}

\begin{document}

\begin{titlepage}

\thispagestyle{empty}

\vspace*{1cm}

\centerline{\large \bf Dual technicolor with hidden local symmetry}

\vspace{1cm}

\centerline{\sc A.V. Belitsky}

\vspace{10mm}

\centerline{\it Department of Physics, Arizona State University}
\centerline{\it Tempe, AZ 85287-1504, USA}

\vspace{2cm}

\centerline{\bf Abstract}

\vspace{5mm}

We consider a dual description of the technicolor-like gauge theory within the D4/D8
brane configuration with varying confinement and elecroweak symmetry breaking scales.
Constructing an effective truncated model valid below a certain cut-off, we identify the
particle spectrum with Kaluza-Klein modes of the model in a manner consistent with the
hidden local symmetry. Integrating out heavy states, we find that the low-energy action
receives nontrivial corrections stemming from the mixing between Standard Model and heavy
gauge bosons which results in reduction of oblique parameters.

\end{titlepage}

\setcounter{footnote} 0

\newpage

\pagestyle{plain}
\setcounter{page} 1

\noindent {\bf 1. Top-down holographic technicolor.} Unravelling the correct mechanism for
electroweak symmetry breaking (EWSB) is one of the most important problems facing particle physics.
The Standard Model (SM) Higgs boson, endowing gauge bosons and fermions with masses, introduces
at the same time the gauge hierarchy problem between the electroweak and Planck scales in the
theory. This makes the Higgs mass unstable under radiative effects thus making the construct
undesirable theoretically. An intriguing alternatives explored for quite some time, which bypasses
this problem, is the dynamical EWSB by condensation of new kinds of fermions coupled by a strongly
interacting gauge sector \cite{WeiSus75,HilSim04}. These technicolor models are hard to tackle
theoretically, however, and attempts to use scaled-up versions of QCD, where the phenomenon of
chiral symmetry breaking mimics the one of EWSB, led to predictions inconsistent with precision
measurements \cite{HolTer90,PesTak90}. Thus the new strongly coupled sector, if realized in Nature,
is not QCD-like. Versions of walking \cite{Hol85}, conformal \cite{LutOku04}, etc., technicolor
were proposed but all plagued by the same calculability problem. Therefore, it appears that the
quest for a consistent model of technicolor is not over.

Recent advances in gauge/string dualities open a window for construction of calculable models
of dynamical EWSB. In this framework, a way to analyze strong coupling region of gauge theory
via weakly-coupled gravitational description enables one to treat non-perturbative dynamics of
technicolor theory in a perturbative fashion. Applied to the problem at hand, the holographic
dual description explores the regime where the probe branes describe the action below the scale
of techniquark condensation. The D-brane configuration in flat space background of type IIA string
theory which realizes a technicolor scenario as an effective theory on the D-branes is based on
the embedding of $N_f$ D8-$\overline{\rm D8}$ branes \cite{SakSug04} in the background of $N$ D4
branes \cite{Wit98} and intersecting in four-dimensional space-time. The gauge fields on the D8
($\overline{\rm D8}$) branes possess $SU_L (N_f )$ ($SU_R (N_f )$) gauge symmetry while techniquark
strings live on D4-D8 (D4-$\overline{\rm D8}$) intersections. This construction is dual to a
confining gauge theory with massless fermions and realizes the $SU_L(N_f) \times SU_R (N_f)$
non-Abelian global symmetry broken down to the diagonal subgroup $SU_D (N_f)$. In its original
incarnation, it was used for modelling chiral symmetry breaking in QCD. Its adaptation to models
of holographic technicolor is achieved by setting $N_f = 2$ and interpreting the gauge symmetry in
the bulk as a weakly gauged symmetry of the four-dimensional gauge theory,  as was done earlier
in Refs.\ \cite{CarErlShe07,HirYos07,MinSon09}. This ensures the custodial global $SU(2)$ symmetry
for the boundary gauge theory which protects isospin observables from receiving large corrections
\cite{Aga03}. To uncover the Standard Model at low energies, the $SU_R (2)$ group is broken down to
$U (1)$ on the boundary by an adjoint scalar living on $\overline{\rm D8}$ with a divergent vacuum
expectation value, which translates to the Dirichlet boundary condition imposed on the corresponding
components the gauge fields.

While attempting to resolve the gauge hierarchy problem, these higher-dimensional theories also
address the issue of unitarity of scattering amplitudes of massive gauge bosons, which in the SM
is cured by the Higgs boson, by means of the exchange of towers of massive Kaluza-Klein (KK) gauge
bosons \cite{Uni01,CsaGroMurPilTer03}.

The starting point for the analysis is the near-horizon geometry of $N$ coincident D4-branes
\cite{Wit98},
\be
d s^2 = - \left( \frac{u}{R} \right)^{3/2}
\left[ \eta_{\mu\nu} d x^\mu d x^\nu - f (u) d \tau^2 \right]
+
\left( \frac{R}{u} \right)^{3/2}
\left[ f^{-1} (u) d u^2 + u^2 d {\mit\Omega}_4^2 \right]
\, ,
\ee
with $R = \pi g_s N (\alpha^\prime)^{3/2}$ and $f (u) = \left( 1 - \frac{u_K^3}{u^3} \right)$.
They extend into the four-dimensional Minkowski space-time with the (mostly negative) metric tensor
$\eta_{\mu\nu}$ and are compactified on a circle in the $\tau$-direction with radius $m_K^{-1}$ in
order to avoid conical singularities. Introducing $N_f \ll N$ D8 branes into this background, these
can be treated in the probe approximation and their embedding is determined by the $u = u (\tau)$
profile, a solution to \cite{AhaSonYan06}
\be
(u^\prime)^2
=
\left( \frac{u}{R} \right)^3
\frac{f (u)^2 \left[ u^8 f (u) - u_0^8 f (u_0) \right]}{u_0^8 f (u_0)}
\, .
\ee
With $u^\prime$ vanishing at $u_0$, the D8 and $\overline{\rm D8}$ branes are smoothly connected
at $u_0 \leq u_K$, admitting a U-shaped form. This configuration geometrically realizes the
dynamical $SU_L (N_f) \times SU_R (N_f)$ symmetry breaking in the dual gauge theory. To distinguish
the D8 and $\overline{\rm D8}$ branches of the resulting solution, a new variable is particularly
convenient
\be
u^3 = u_0^3 (1 + z^2)
\, ,
\ee
which goes along $\overline{\rm D8}$ branch for $- z_R \leq z \leq 0$ and D8 for $0 \leq z \leq z_L$
with cut-offs $z_{L,R}$ reflecting the finite volumes of the electroweak branes. Then the probe
D8-brane Dirac-Born-Infeld (DBI) action encodes the EWSB which endows the SM gauge bosons, identified
with the lowest modes in the KK expansions of the brane gauge field, and other technimesons with masses.
Ignoring entirely the towers of modes with nonvanishing angular momentum on $S^4$ as well as all gauge
fields along these transverse directions as being heavier than KK states with respect to the compact
$z$-direction, we can rewrite the quadratic part of the five-dimensional DBI action in the unitary
$A_z = 0$ gauge as
\be
\mathcal{S} = - \frac{1}{g_5^2} {\rm tr} \, \int d^4 x \, d z
\left\{
\frac12 w_1 (z) F_{\mu\nu}^2
-
\varepsilon^{-1/3} m_K^2 w_2 (z) \left( \partial_z A_\mu \right)^2
\right\}
\, ,
\ee
where $\varepsilon \equiv (u_K/u_0)^3 \leq 1$ and
\be
g_5^{-2} = \ft{2}{3} u_0^{1/2} T_8 V_4 g_s^{-1} (2 \pi \alpha^\prime)^2 R^{9/2}
\, , \qquad
m_K^2 = \ft{9}{4} \frac{u_K}{R^3}
\ee
are the five-dimensional coupling and the compactifiction scale, respectively. The weights
stemming from the warped metric are
\be
w_1 (z) = (1 + z^2)^{2/3} w_2^{-1} (z) = (1 + z^2)^{-1/3}
\left(
1 + (1 - \varepsilon) \frac{(1 + z^2)^{5/3} - 1}{z^2 (1 + z^2)^{5/3}}
\right)^{-1/2}
\, .
\ee
The modes diagonalizing the above action obey the eigenvalue equation
\be
\label{EigenModeEquation}
\partial_z \left( w_2 (z) \partial_z \psi_n \right) + \varepsilon^{1/3} \lambda_n w_1 (z) \psi_n = 0
\, .
\ee
To have dynamical fields in the ultraviolet on the D8 and $\overline{\rm D8}$, the boundary
conditions are imposed as follows:
\be
\label{BCs}
\partial_z A^\pm_\mu (z_L) = 0 \, , \qquad \partial_z  A^3_\mu (z_L) = 0
\, , \qquad
A^\pm_\mu (- z_R) = 0 \, , \qquad \partial_z  A^3_\mu (- z_R) = 0
\, .
\ee

There are different ways to identify field content of the model with physical particle spectra. In
the present note we will employ a residual gauge freedom of the five-dimensional gauge fields
describing the fluctuation of the D-branes to identify (axial) vector mesons differently to the
Callan-Coleman-Wess-Zumino (CCWZ) formulation \cite{CalColWesZum69}. This hidden local symmetry (HLS)
\cite{BanKugYam85,CasCurDomGat87} introduces a kinetic mixing among the light gauge bosons and their
KK excitations. Integrating out the heavy modes from the spectrum will define a low-energy theory with
non-Standard Model interactions encoded in the renormalization factors. Coupling SM fermions to
five-dimensional gauge bosons, we compute the oblique parameters \cite{PesTak90,BurGodKonLonMak93}
in this model and demonstrate that KK axial towers tend to reduce the value of the $S$ parameter,
which is of order one in typical technicolor models.

\noindent {\bf 2. CCWZ vs. HLS.} Within the CCWZ approach the fields in the KK decomposition
\be
A_\mu^a (x, z) = \sum_{n \geq 0} \psi^a_n (z) \, A_\mu^{a(n)} (x)
\, ,
\ee
with $a = \pm, 3$, corresponds to observed particles. Here the lowest components are identified with
the photon $A^{3(0)}_\mu = B^0_\mu$ and neutral $A^{3(1)}_\mu = Z^0_\mu$ and charged $A^{\pm(0)}_\mu
= W^\pm_\mu$ gauge bosons, respectively, while the rest with heavier mass modes. Then, the diagonalized
quadratic part of the DBI action written in terms of mass eigenstates reads
\be
S = - \frac{1}{g_5^2} \sum_{a = 0, \pm} \sum_{n \geq 0} N^a_n \int d^4 x
\left\{
\frac{1}{4} \left( F_{\mu\nu}^{a (n)} \right)^2 - \frac{1}{2} m_{n, a}^2
\left( A_\mu^{a (n)} \right)^2
\right\}
\, ,
\ee
where the masses and normalization constants are
\be
m_{n, a}^2 = \lambda_{n, a} m_K^2
\, , \qquad
N_n = \int_{- z_R}^{z_L} dz \, w_1 (z) \left( \psi_n (z) \right)^2
\, .
\ee

Notice however that in the unitary gauge, the gauge field $A^a_\mu (x, z)$ may transform under
a residual gauge transformation $h(x) = U (x, 0)$, independent of $z$ variable, which leaves
the condition $A_z = 0$ invariant. However, while Neumann boundary conditions are consistent
with it, the Dirichlet boundary conditions acquire a nontrivial right-hand side for the gauge
transformed variables. This is a reflection of the well-known fact that the physical gauges,
like axial, light-like, etc., require boundary conditions imposed on fields to fix the gauge
symmetry completely and get of rid of residual degeneracies in gauge theories which prohibit
their consistent quantization. For the present setup, this implies that while the third
isovector component $A^3_\mu$ of the gauge field allows for hidden local symmetry transformations
\cite{BanKugYam85,CasCurDomGat87}, the $A^\pm_\mu$ ones do not, i.e.,
\ba
\label{GaugeTransform5D}
A^\pm_\mu (x, z)
\!\!\!&\to&\!\!\!
h (x) A^\pm_\mu (x, z) h^\dagger (x)
\, \nonumber\\
A^3_\mu (x, z)
\!\!\!&\to&\!\!\!
h (x) A^3_\mu (x, z) h^\dagger (x) + i h (x) \partial_\mu h^\dagger (x)
\, .
\ea
The general KK decomposition is then
\ba
\label{HLSdecomposition}
A^\pm_\mu (x, z)
\!\!\!&=&\!\!\!
\sum_{n \geq 0}
W_\mu^{\pm (n)} (x) \phi^\pm_n (z)
\, , \\
A^3_\mu (x, z)
\!\!\!&=&\!\!\!
\sum_{n \geq 0}
\left\{
L_\mu^{(n)} (x) \phi_{L, n} (z) + R_\mu^{(n)} (x) \phi_{R, n} (z)
\right\}
\, ,
\ea
where the neutral field is decomposed into left and right modes reflecting the symmetry of the
boundary conditions imposed on it and emphasizing the fact that, contrary to the charged vector,
either of the boundaries will yield light dynamical modes identified with the photon and $Z$-boson.
As a consequence, while the charged modes transform homogeneously under the HLS and we can identify
\be
\label{ChargedEigenFunctions}
\phi^\pm_n = \psi^\pm_n
\, ,
\ee
where $\psi_n^\pm$ are solutions to Eq.\ \ref{EigenModeEquation}, the neutral bosons are
inhomogeneous,
\be
\label{GaugeTransform4DKK}
\left( L_\mu^{(n)} (x), R_\mu^{(n)} (x) \right)
\to
h (x) \left( L_\mu^{(n)} (x), R_\mu^{(n)} (x) \right) h^\dagger (x)
+
i h (x) \partial_\mu h^\dagger (x)
\, .
\ee
The consistency between Eqs.\ (\ref{GaugeTransform5D}) and (\ref{GaugeTransform4DKK}) implies
that
\be
\sum_{\alpha = L, R} \sum_{n \geq 0} \phi_{\alpha, n} (z) = 1
\, ,
\ee
where the modes $\phi_{\alpha, n}$ are not the eigenmodes of the equation of motion but rather
their linear combinations which we are about to construct. First, to resolve the consistency
condition we introduce the following linear combinations
\be
\phi_\alpha^k = \psi_\alpha^k - \psi_\alpha^{k + 1}
\, , \qquad\mbox{for}\qquad k = 0, \dots, M - 1
\, , \qquad
\phi_\alpha^{M} = \psi_\alpha^{M}
\, ,
\ee
where we truncated the series at some total number of (left and right) modes $2 M + 2$,
understanding that we are dealing with a low-energy effective theory. The above condition
suggests that the sum of the lowest two modes gives the zero mode of the eigenvalue
problem (\ref{EigenModeEquation}),
\be
\psi_0 (z) = \psi_{L, 0} (z) + \psi_{R, 0} (z) = 1
\, .
\ee
Next, the fifth-dimension wave functions of the photon and $Z$-boson as well as their
KK excitations are related to the modes resolving the consistency condition via the
following transformation
\ba
\left(
\begin{array}{l}
\psi_{R, k}
\\
\psi_{L, k}
\end{array}
\right)
=
\left(
\begin{array}{lr}
\cos^2 \theta_k & \sin \theta_k \, \cos \theta_k
\\
\sin^2 \theta_k & - \sin \theta_k \, \cos \theta_k
\end{array}
\right)
\left(
\begin{array}{l}
\psi_{2 k}
\\
\psi_{2 k + 1}
\end{array}
\right)
\, ,
\ea
whose unusual form is used to restore the conventional normalization of the gauge fields
in the decomposition of the left and right vector fields in  Eqs.\ (\ref{VectorL}) and
(\ref{VectorR}) below. Here the wave functions $\psi_n$ on the right-hand side are indeed
the eigenstates of Eq.\ (\ref{EigenModeEquation}). The lowest mixing angle $\theta_0$ has
to be chosen to coincide with the Weinberg angle $\theta_W$. Finally, the gauge eigenstates
$B_\mu^{(n)}$ and $Z_\mu^{(n)}$, which transform in- and homogeneously, respectively, with
respect to the HLS (\ref{GaugeTransform4DKK}),
\be
B_\mu^{(n)} (x) \to h (x) B_\mu^{(n)} (x) h^\dagger (x) + i h (x) \partial_\mu h^\dagger (x)
\, , \qquad
Z_\mu^{(n)} (x) \to h (x) Z_\mu^{(n)} (x) h^\dagger (x)
\, ,
\ee
arise in the decomposition of the left and right gauge fields as
\ba
\label{VectorL}
L_\mu^{(n)}
&=&
B_\mu^{(n)} - \sum_{k = 1}^n
\left(
\sin^2 \theta_k \, \cot \theta_{k - 1} - \cos^2 \theta_k \, \tan \theta_{k - 1}
\right)
Z_\mu^{(k - 1)}
-
\cot \theta_n \, Z_\mu^{(n)}
\, , \\
\label{VectorR}
R_\mu^{(n)}
&=&
B_\mu^{(n)} - \sum_{k = 1}^n
\left(
\sin^2 \theta_k \, \cot \theta_{k - 1} - \cos^2 \theta_k \, \tan \theta_{k - 1}
\right)
Z_\mu^{(k - 1)}
+
\tan \theta_n \, Z_\mu^{(n)}
\, .
\ea
Here a linear combination of homogeneous $Z^{(k < n)}_\mu$-fields was absorbed into the KK vector
modes $B^{(n)}_\mu$ so as to eliminate the mixing between the vector and axial modes.

Substituting these results into the decomposition (\ref{HLSdecomposition}), we obtain the final
KK expansion
\ba
A_\mu^3 (x, z)
\!\!\!&=&\!\!\!
B^{(0)}_\mu (x) \psi_0 (z) + Z^{(0)}_\mu (x) \psi_1 (z)
\\
&+&\!\!\!
\sum_{k = 1}^M \left[ B_\mu^{(k)} (x) - B_\mu^{(k - 1)} (x) \right] \psi_{2 k} (z)
+
\sum_{k = 1}^M \left[ Z_\mu^{(k)} (x) - f_k Z_\mu^{(k - 1)} (x) \right] \psi_{2 k + 1} (z)
\, , \nonumber
\ea
with self-obvious HLS properties. From the above definitions, it follows that $\psi_0 (z) = 1$ and
\be
f_k = \frac{\sin \theta_k \, \cos \theta_k}{\sin \theta_{k - 1} \, \cos \theta_{k - 1}}
\, ,
\ee
with $f_0 = 1$. The four-dimensional neutral gauge boson action then splits into two orthogonal
sectors and reads
\be
\mathcal{S}_{n} = - \frac{1}{2 g_5^2} \int \frac{d^4 p}{(2 \pi)^4}
\sum_{j, k = 0}^M
\left\{
B_\mu^{(j)} (p) {\mit\cal V}_{\mu\nu}^{jk} (p) B_\mu^{(k)} (- p)
+
Z_\mu^{(j)} (p) {\mit\cal A}_{\mu\nu}^{jk} (p) Z_\mu^{(k)} (- p)
\right\}
\, ,
\ee
where
\ba
\label{Vmatrix}
{\mit\cal V}_{\mu\nu}^{jk} (p)
\!\!\!&=&\!\!\!
P_{\mu\nu} (p)
\left\{
(\alpha_j + \alpha_{j + 1}) \delta_{jk} - \alpha_j \delta_{j - 1, k} - \alpha_{j + 1} \delta_{j + 1, k}
\right\}
\nonumber\\
&-&\!\!\!
\eta_{\mu\nu}
\left\{
(\mu_j^2 + \mu^2_{j + 1}) \delta_{jk} - \mu^2_j \delta_{j - 1, k} - \mu^2_{j + 1} \delta_{j + 1, k}
\right\}
\, , \\[2mm]
\label{Amatrix}
{\mit\cal A}_{\mu\nu}^{jk} (p)
\!\!\!&=&\!\!\!
P_{\mu\nu} (p)
\left\{
(\beta_j + f_{j + 1}^2 \beta_{j + 1}) \delta_{jk}
- \beta_j f_j \delta_{j - 1, k} - \beta_{j + 1} f_{j + 1} \delta_{j + 1, k}
\right\}
\nonumber\\
&-&\!\!\!
\eta_{\mu\nu}
\left\{
(\nu_j^2 + f_{j + 1}^2 \nu^2_{j + 1}) \delta_{jk} - f_j \nu^2_j \delta_{j - 1, k}
-
f_{j + 1} \nu^2_{j + 1} \delta_{j + 1, k}
\right\}
\, ,
\ea
with the kinetic projection operator being
\be
P_{\mu\nu} (p) =
p^2 \eta_{\mu\nu} - p_\mu p_\nu
\, .
\ee
The mixing matrices (\ref{Vmatrix}) and (\ref{Amatrix}) are defined in terms of the
normalization constants for even $\alpha_k = N^0_{2k}$ and odd $\beta_k = N^0_{2k + 1}$
modes and corresponding mass parameters $\mu^2_k = \alpha_k m_{2 k}^2$ and $\nu^2_k =
\beta_k m_{2k + 1}^2$, respectively. Notice that $\mu_0^2 = 0$. At the same time, the
charged boson action is diagonal from the outset due to identification
(\ref{ChargedEigenFunctions}) driven by the boundary conditions (\ref{BCs}) such that
the lowest component is indeed the physical $W$, however with unconventional normalization
$\gamma_{0, \pm} = N_0^\pm$ of the kinetic term and mass $\mu_{0, \pm}^2 =
\gamma_{0,\pm} m_{0, \pm}^2$.

As a next step, we find the eigenstates of the mass matrix and integrate out all KK modes
yielding an effective Lagrangian for the lightest modes only in each sector. As a result,
the photon $B_\mu$ and $Z$-boson $Z_\mu$ mass eigenstates are given by linear combinations
of the gauge eigenstates $B^{(k)}_\mu$ and $Z^{(k)}_\mu$,
\be
B_\mu
=
\frac{1}{M + 1} \sum_{k = 0}^M B_\mu^{(k)}
\, , \qquad
Z_\mu
\simeq
\frac{1}{D_M} \sum_{k = 0}^M \left( \prod_{n = 1}^k f_n \right) Z_\mu^{(k)}
\, ,
\ee
where the last expression is quoted to $\mathcal{O} (\delta)$ accuracy only with $\delta =
\max \nu_j^2/\nu_{k > j}^2$. Here we introduced the following notation for the function
of mixing angles
\be
D_k \equiv \sum_{m = 0}^k \prod_{n = 0}^m f_n^2
\, .
\ee

The ratio of the five-dimensional cut-offs $z_R/z_L$ is correlated with the value of the
model's Weinberg angle $\theta_0$. The lowest states in the KK towers, i.e., photon, $Z$ and
$W$, are ultralight (UL) and separated hierarchically from the rest of the KK spectrum
by a gap typical to gap-like metrics \cite{HirSan07,CuiGheWel09},
\be
\frac{m_{\rm UL}}{m_{K}} \sim \left( \int^{z_{L,R}} dz \, w_1 (z) \right)^{-1/2}
\, .
\ee

Summarizing our findings, the low-energy action including the charged-boson sector then reads
\ba
\mathcal{S} = - \frac{1}{2 g_5^2}
\int \frac{d^4 p}{(2 \pi)^4}
\bigg\{
P_{\mu\nu} (p)
\big[
B_\mu (p) B_\nu (- p) K_B
\!\!\!&+&\!\!\!
Z_\mu (p) Z_\nu (- p) K_Z
+
2
W_\mu (p) W^\ast_\mu (- p) K_W
\big]
\nonumber\\
&-&\!\!\!
\nu_0^2 Z_\mu (p) Z_\mu (- p)
-
2 \mu_{0, \pm}^2 W_\mu (p) W_\mu^\ast (- p)
\bigg\}
\, , \qquad
\ea
where the photon remains massless and the kinetic terms are
\ba
K_B \!\!\!&=&\!\!\! \alpha_0 + \mathcal{O} (p^2)
\, , \\
K_Z \!\!\!&=&\!\!\! \beta_0
 \left(
1 - 2 \sum_{k = 1}^M
\frac{\nu_0^2}{\nu_k^2}
\left( 1 - \frac{D_{k - 1}}{D_M} \right)^2
\prod_{m = 1}^k f_m^{-2}
+
\mathcal{O} (\delta^2)
\right)
+ \mathcal{O} (p^2)
\, , \\
K_W \!\!\!&=&\!\!\! \gamma_{0, \pm}
\, .
\ea
Here we did not display $\mathcal{O} (p^2)$ terms since they do not affect Peskin-Takeuchi
parameters.

\VerbatimFootnotes

\noindent {\bf 3. Spectrum and oblique parameters.} The mass spectrum emerges from the
eigenvalue equation and can be computed numerically\footnote{
Here is a simple routine written in Mathematica to solve the boundary value problem Eq.\
(\ref{EigenModeEquation}) by reducing it to an initial value problem with a variable
boundary condition that the software environment can handle with built-in commands
\begin{verbatim}
sys[s_, k_]:=
{D[w2[z]*F1[z], z]+eps^(1/3)*k*w1[z]*F[z]==0, D[F[z], z]==F1[z], F[zR]==VR, F1[zR]==s};
solIV[s_?NumericQ, k_?NumericQ]:={F, F1} /.NDSolve[sys[s, k], {F, F1}, {z, zL, zR}][[1]];
F1atzL[s_?NumericQ, k_]:=solIV[s, k][[2]][zL];
sFromzL[k_]:=FindRoot[F1atzL[s, k]==VL, {s, s0min, s0max}, MaxIterations -> 100];
solBV[k_?NumericQ, z_?NumericQ]:=solIV[sFromzL[k][[1, 2]], k][[1]][z];
\end{verbatim}
While this program is used for the calculation of the $W$-tower KK eigenspectrum, its
modification to accommodate Neumann-Neumann boundary conditions is self-obvious.
}.
To calculate precision electroweak corrections we have to find a scheme in which all
corrections will be oblique, i.e., the effective four-dimensional Lagrangian describing
the coupled gauge boson-fermion system only gets corrections in the gauge boson sector,
but none the fermionic sector. There are a couple of ways to achieve this. Either, with the
bulk gauge kinetic terms normalized by $g_5^{-2}$, a very simple convention is to set the
gauge boson wave functions to one at the location of the fermion \cite{CacCsaGroTer04}
and require that the couplings of localized fermions reproduce exactly the leading order
relations between them. Or, by choosing the canonical normalization for the four-dimensional
gauge boson kinetic terms, channel all new physics into gauge-fermion couplings
\cite{BurGodKonLonMak93}. In this work, we choose the latter.

The coupling of gauge fields through the covariant derivative to the SM fermions, localized
in the ultraviolet, has the structure\footnote{Obviously, since we are allowing for KKs to
couple to light fermions, the model induces non-oblique corrections as well.}
\be
\bar\lambda (x) \gamma^\mu \ft12 \sigma_a A_\mu^a (x, z_L) \lambda (x)
\, ,
\ee
where the strength of the interaction is included into the five-dimensional gauge field as
is exhibited by the normalization of its kinetic term. Integrating out the heavy KK modes,
the gauge fields are effectively replaced by $A_\mu^\pm (x, z_L) \to W_\mu^\pm (x) \psi_0^\pm
(z_L)$ and
\ba
&&
\!\!\!\!\!\!\!\!\!
A_\mu^3 (x, z_L)
\to
B_\mu (x) \psi_0 (z_L)
\\
&&\qquad
+ \,
Z_\mu (x) \psi_1 (z_L)
\bigg\{
1
-
\sum_{k = 1}^M
\frac{\nu_0^2}{\nu_k^2}
\left( 1 - \frac{D_{k - 1}}{D_M} \right)
\prod_{m = 1}^k f_m^{-1}
\bigg[
\left( 1 - \frac{D_{k - 1}}{D_M} \right)
\prod_{m = 1}^k f_m^{-1}
-
\frac{\psi_{2k + 1} (z_L)}{\psi_1 (z_L)}
\bigg]
\bigg\}
\, , \nonumber
\ea
where the result is valid to order $\delta^2$ and we ignored four-fermion contributions.

Since the HLS requires $\psi_0 = 1$, and the electromagnetic group remains unbroken, we
can impose the standard normalization on the photon kinetic term. This eliminates oblique
corrections from the photon sector and yields the relation
\be
g_5^2 = \alpha_0 \, e^2
\, ,
\ee
where $e = \sqrt{4 \pi \alpha_{\rm em}} = g \sin \theta_W$ is the electric charge, expressed in
terms of the $SU(2)$ coupling $g$. Rescaling the photon field, $B_\mu \to e B_\mu$, the strength
of the interaction migrates to the photon-fermion terms. The standard normalization for the SM
massive bosons is achieved by rescaling the $Z$- and $W$-fields $Z_\mu \to g_5 K_Z^{-1/2} Z_\mu$,
$W_\mu \to g_5 K_W^{-1/2} W_\mu$, such that the bosons masses and boson-fermion interaction
couplings read
\be
m_W = m_{0, \pm}
\, , \qquad
m_Z
\simeq
m_1 \left(
1 + \sum_{k = 1}^M
\frac{\nu_0^2}{\nu_k^2}
\left( 1 - \frac{D_{k - 1}}{D_M} \right)^2
\prod_{m = 1}^k f_m^{-2}
\right)
\ee
and
\ba
g_{cc}
\!\!\!&=&\!\!\!
e \, \psi^\pm_1 (z_L)  \left( \frac{\alpha_0}{\gamma_{0, \pm}} \right)^{1/2}
\, , \nonumber\\
g_{nc}
\!\!\!&\simeq&\!\!\!
e \, \psi_1 (z_L) \left( \frac{\alpha_0}{\beta_0} \right)^{1/2}
\bigg(
1
+
\sum_{k = 1}^M \frac{\psi_{2k + 1} (z_L)}{\psi_1 (z_L)}
\frac{\nu_0^2}{\nu_k^2}
\left( 1 - \frac{D_{k - 1}}{D_M} \right)
\prod_{m = 1}^k f_m^{-1}
\bigg)
\, ,
\label{NeutralCurrentCoupling}
\ea
respectively.

\begin{table}[t]
\scriptsize
\begin{center}
\begin{tabular}{|c|c||c|c|c|c|c|c|c|c|c|c|}
\hline $\varepsilon$ & $m_K$ & $\nu_1$ & $\nu_2$ & $\nu_3$ & $\nu_4$ & $\nu_5$
& $\nu_6$ & $\nu_7$ & $\nu_8$ & $\nu_9$ & $\nu_{10}$
\\ [1mm]
\hline $1$ & $3593.6$ & $4512.7$ & $7681.4$ & $10812.6$ & $13932.9$ & $17048.2$
& $20160.9$ & $23272.0$ & $26382.1$ & $29491.5$ & $34154.7$
\\ [1mm]
\hline $\ft{1}{64}$ & $1597.3$ & $4454.3$ & $7638.6$ & $10762.3$ & $13867.9$ & $16966.8$
& $20062.8$ & $23157.2$ & $26250.8$ & $29343.7$ & $32436.2$
\\ [1mm]
\hline $\ft{1}{512}$ & $1128.2$ & $4453.6$ & $7638.0$ & $10761.6$ & $13867.0$ & $16965.6$
& $20061.4$ & $23155.6$ & $26248.9$ & $29341.6$ & $32433.9$
\\ [1mm]
\hline
\end{tabular}
\end{center}
\caption{The KK scales and masses (in GeV) of the $Z$-boson KK tower for $z_L = 10^7$ and different
values of $\varepsilon$.}
\label{NeutralMasses}
\end{table}%

The oblique corrections are obtained by inverting the matrix of lepton-boson gauge coupling
constants and the $W$ mass \cite{BurGodKonLonMak93}, yielding the result
\be
\left(
\begin{array}{c}
\alpha_{\rm em} (S - 2 c_W^2 T) \\
\alpha_{\rm em} U \\
\Delta
\end{array}
\right)
=
\left(
\begin{array}{ccc}
4 c_W^2 (c_W^2 - s_W^2) & - 4 c_W^2 & 2 c_W^2 \\
8 s_W^2           & - 8 s_W^2 & 0       \\
0                 & 2         & - 1
\end{array}
\right)
\left(
\begin{array}{c}
1 - \frac{s_W}{e c_W} g_{nc} \\
1 - \frac{s_W}{e} g_{cc} \\
1 - \frac{m_W^2}{c_W^2 m_Z^2}
\end{array}
\right)
\, .
\ee
Since we are not introducing an additional $U(1)$ coupling (to avoid additional model
dependence) to obtain correct hypercharges, we will not be able to separate $S$ and $T$
parameters in the combination $S - 2 c_W^2 T$. However, the fact that the model enjoys
the custodial symmetry in the bulk implies that the $T$ parameter is small and thus the
above combination is dominated by $S$.

\begin{figure}[t]
\begin{center}
\mbox{
\begin{picture}(0,230)(230,0)
\put(-10,0){\insertfig{7.65}{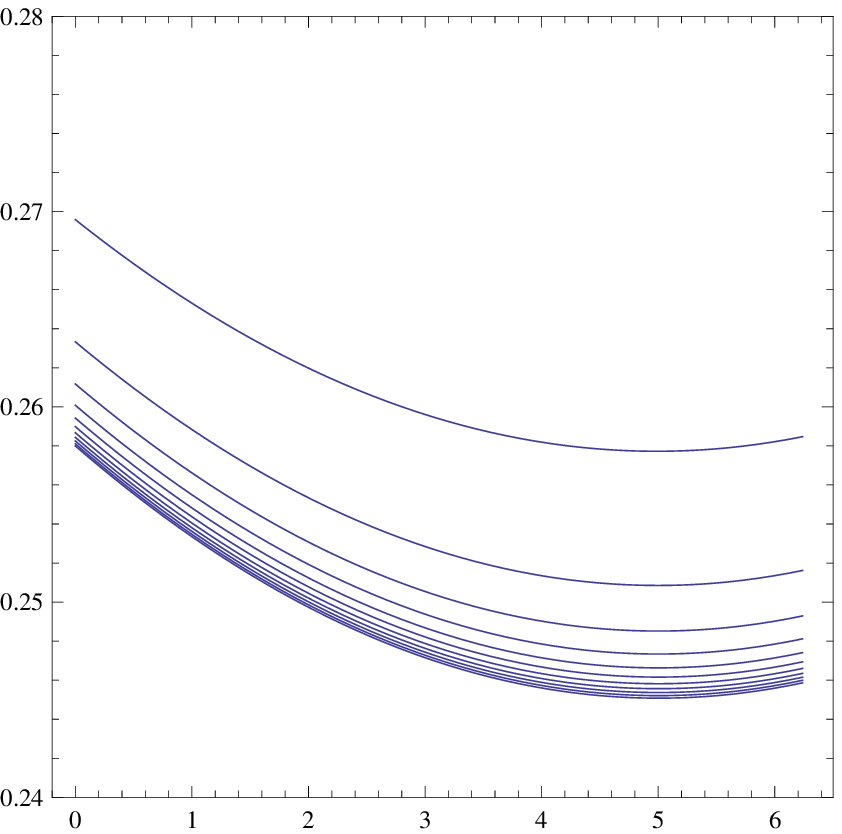}}
\put(84,197){$(S - 2 c_W^2 T) \ \mbox{vs.} \ \log \varepsilon^{-1}$}
\put(180,106){$\scriptstyle{M = 0}$}
\put(180,72){$\scriptstyle{M = 1}$}
\put(180,60){$\scriptstyle{M = 2}$}
\put(176,28){$\scriptstyle{M = 10}$}
\put(250,0){\insertfig{7.65}{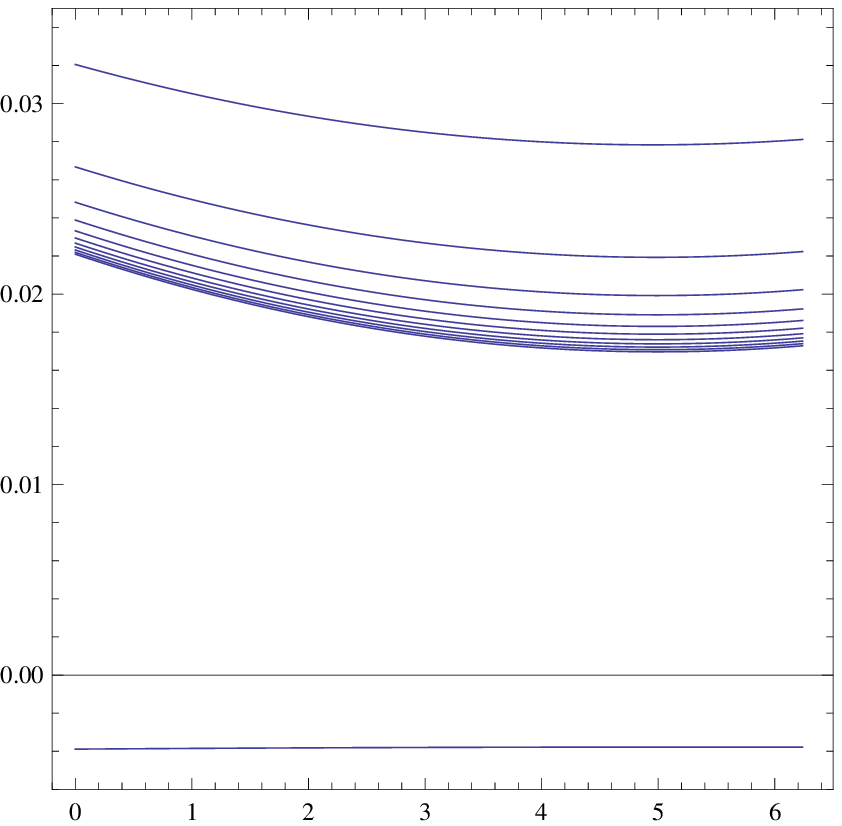}}
\put(395,45){$U \ \mbox{vs.} \ \log \varepsilon^{-1}$}
\put(395,26){$\Delta \ \mbox{vs.} \ \log \varepsilon^{-1}$}
\put(440,184){$\scriptstyle{M = 0}$}
\put(440,153){$\scriptstyle{M = 1}$}
\put(440,142){$\scriptstyle{M = 2}$}
\put(436,114){$\scriptstyle{M = 10}$}
\end{picture}
}
\end{center}
\caption{ \label{Oblique} The oblique parameters as a function of $\log \varepsilon^{-1}$ for
increasing number of KK modes $M$ for the cut-off $z_L = 10^7$.}
\end{figure}

Below, we choose unit normalizations for all wave functions, i.e., $N^a_{n} = 1$, except for
$\alpha_0$ which cannot be altered due to its constrained form by HLS. The value of $\psi_W$
at $z_L$ is very weakly dependent on $z_R$ which, to the accuracy that we worked with, could be
ignored. Since the overall phase of the KK wave functions is not automatically fixed, we choose
the same sign for their asymptotic values as for the ultralight modes, i.e., $\psi_{2k + 1}
(z_L)/\psi_1 (z_L) > 0$. The values of the KK wave functions, localized in the vicinity of $z = 0$,
are suppressed stronger at the cut-off $z_L$ compared to the ultralight modes, by a factor
$\mathcal{O} (10^{-1})$. The holographic description, which we advocated here, is valid in the
strong coupling regime on the gauge theory side, i.e., when the 't Hooft coupling $\lambda \equiv
g^2_{\rm YM} N \gg 1$. Its value can be estimated from the following equation
\be
\lambda = \frac{27 \pi^2}{N \alpha_{\rm em} \, \alpha_0}
\, ,
\ee
which imposes an upper limit on the cut-off parameters $z_L$ for reliable applicability of
holography.

We analyzed the mass spectra and oblique corrections as functions of the cut-offs, $z_{L,R}$
and parameter $\varepsilon$ separating the confinement and electroweak symmetry breaking scale
fixing two parameters of the model to the $Z$ mass $m_Z$ and the fine structure constant $\alpha_{\rm
em}$ at the $Z$ pole. For $z_{L,R} < 10^6$, we found that the $S$ parameter is greater than one
and this part of the parameter space is excluded by precision electroweak measurements. For
$z_L = 10^6$, we varied the right cut-off in the interval $30 \leq z_R/z_L \leq 40$. Ignoring the
contribution of KK towers, i.e., setting $M = 0$ in Eq.\ (\ref{NeutralCurrentCoupling}), one finds
that the mixing angle $s^2_0 = 1 - \mu_{0,\pm}^2/\nu_0^2$, the KK scale $m_K$ (in GeV) and the
$S$-parameter vary in the intervals
\baa
&&\!\!\!\!
0.24355 \leq s^2_0 \leq 0.22626 \, , \quad
2421.06 \leq m_K \leq 2448.57 \, , \quad
0.719 \leq (S - 2 c_W^2 T) \leq 0.623 \, , \\
&&\!\!\!\!
0.24328 \leq s^2_0 \leq 0.22599 \, , \quad
1075.79 \leq m_K \leq 1088.01 \, , \quad
0.687 \leq (S - 2 c_W^2 T) \leq 0.596 \, , \\
&&\!\!\!\!
0.24327 \leq s^2_0 \leq 0.22599 \, , \quad
759.864 \leq m_K \leq 768.494 \, , \quad
0.683 \leq (S - 2 c_W^2 T) \leq 0.601 \, ,
\eaa
for $\varepsilon = 1, \ft{1}{64}, \ft{1}{512}$, respectively. For $z_L = 10^7$, we fixed $s_0$ to
the experimental value $s^2_0 = s_W^2 = 0.23108$, such that the tree-level $\rho = 1$, which
corresponds to the choices $z_R/z_L = 36.835, \, 36.772, \, 36.771$ for $\varepsilon = 1, \,
\ft{1}{64}, \, \ft{1}{512}$, respectively. The resulting neutral KK mass spectra are shown in the
Table \ref{NeutralMasses} and the oblique corrections, with the lowest 10 KK modes included with
equal mixing angles, implying that $f_k \simeq 1$, are displayed in Figure \ref{Oblique}. One
finds that in this regime the 't Hooft coupling is moderate $\lambda \simeq 13/N$ (with a very
weak dependence on $\varepsilon$) while higher values of $z_L$, though yield smaller $S$ compatible
with precision electroweak measurements, start falling outside of the strong-coupling regime. While
the contribution of KKs into the oblique corrections definitely reduces the uncorrected $S$-parameter,
it is still large (for $z_L = 10^7$) compared to the experimental values. Though direct comparison
of the two values is not straightforward (see, e.g., \cite{CuiGheWel09} for the most recent discussion)
since the subtraction of the Higgs boson and accompanying vector boson loops was not done, a mechanism
for further reduction of the Peskin-Takeuchi parameter can be achieved by coupling the localized
fermions to the gauge bosons at $z \ll z_L$ where the KK wave functions are larger or by using
delocalized fermions in the bulk \cite{CacCsaGroTer05}. Yet understanding of $1/N$-corrections in
the gauge dual can be done by means of incorporation of the meson one-loop effects stemming from
the DBI actions. Due to HLS the power counting is well-defined in this framework \cite{HarMatYam06}.
These questions will be addressed elsewhere.

\vspace{5mm}

\noindent This work was supported by the U.S. National Science Foundation under grant no.\
PHY-0757394. We would like to thank Josh Erlich for a number of stimulating discussions,
careful reading of the manuscript and instructive comments.




\begin{thebibliography}{99}
%
\bibitem{WeiSus75}
S. Weinberg,
Phys. Rev. D 13 (1976) 974;
%
Phys. Rev. D 19 (1979) 1277; \\
%
L. Susskind,
Phys. Rev. D 20 (1979) 2619.
%
\bibitem{HilSim04}
K. Lane,
{\sl Two lectures on technicolor},
arXiv:hep-ph/0202255; \\
%
C.T.Hill, E.H. Simmons,
Phys. Rept. 381 (2003) 235; (E) 390 (2004) 553; \\
%
F. Sannino,
{\sl Dynamical stabilization of the Fermi scale}, arXiv:0804.0182 [hep-ph];
%
{\sl Conformal dynamics for TeV physics and cosmology},
arXiv:0911.0931 [hep-ph].
%
\bibitem{HolTer90}
B. Holdom, J. Terning,
Phys. Lett. B 247 (1990) 88; \\
M. Golden, L. Randall,
Nucl. Phys. B 361 (1991) 3.
%
\bibitem{PesTak90}
M.E. Peskin, T. Takeuchi,
Phys. Rev. D 46 (1992) 381.
%
\bibitem{Hol85}
B. Holdom,
Phys. Lett. B 150 (1985) 301; \\
%
T.W. Appelquist, D. Karabali, L.C.R. Wijewardhana,
Phys. Rev. Lett. 57 (1986) 957; \\
%
K. Yamawaki, M. Bando, K. Matumoto,
Phys. Rev. Lett. 56 (1986) 1335.
%
\bibitem{LutOku04}
M.A. Luty, T. Okui,
J. High Ener. Phys. 0609 (2006) 070.
%
\bibitem{SakSug04}
T. Sakai, S.~Sugimoto,
Prog. Theor. Phys. 113 (2005) 843.
%
\bibitem{Wit98}
E. Witten,
Adv. Theor. Math. Phys. 2 (1998) 505.
%
\bibitem{CarErlShe07}
C.D. Carone, J. Erlich, M. Sher,
Phys. Rev. D 76 (2007) 015015.
%
\bibitem{HirYos07}
T. Hirayama, K. Yoshioka,
J. High Ener. Phys. 0710 (2007) 002.
%
\bibitem{MinSon09}
O. Mintakevich, J. Sonnenschein,
J. High Ener. Phys. 0907 (2009) 032.
%
\bibitem{Aga03}
K. Agashe, A. Delgado, M. May and R. Sundrum,
J. High Ener. Phys. 0308 (2003) 050.
%
\bibitem{Uni01}
R. S. Chivukula, D. A. Dicus, H. J. He,
Phys. Lett. B 525 (2002) 175.
%
\bibitem{CsaGroMurPilTer03}
C. Csaki, C. Grojean, H. Murayama, L. Pilo, J. Terning,
Phys. Rev. D 69 (2004) 055006; \\
%
C. Csaki, C. Grojean, L. Pilo, J. Terning,
Phys. Rev. Lett. 92 (2004) 101802.
%
\bibitem{AhaSonYan06}
O. Aharony, J. Sonnenschein, S. Yankielowicz,
Annals Phys. 322 (2007) 1420.
%
\bibitem{CalColWesZum69}
C.G. Callan, S.R. Coleman, J. Wess, B. Zumino,
Phys. Rev. 177 (1969) 2247.
%
\bibitem{BanKugYam85}
M. Bando, T. Kugo, K. Yamawaki,
Prog. Theor. Phys. 73 (1985) 1541;
%
Nucl. Phys. B 259 (1985) 493;
%
Phys. Rept. 164 (1988) 217; \\
%
M. Bando, T. Fujiwara, K. Yamawaki,
Prog. Theor. Phys. 79 (1988) 1140; \\
%
M. Harada, K. Yamawaki,
Phys. Rept. 381 (2003) 1.
%
\bibitem{CasCurDomGat87}
R. Casalbuoni, S. De Curtis, D. Dominici, R. Gatto,
Nucl. Phys. B 282 (1987) 235.
%
\bibitem{BurGodKonLonMak93}
C.P. Burgess, S. Godfrey, H. Konig, D. London, I. Maksymyk,
Phys. Rev. D 49 (1994) 6115.
%
\bibitem{HirSan07}
J. Hirn, V. Sanz,
Phys. Rev. D 76 (2007) 044022.
%
\bibitem{CuiGheWel09}
Y. Cui, T. Gherghetta, J.D. Wells,
J. High Ener. Phys. 0911 (2009) 080.
%
\bibitem{CacCsaGroTer04}
C. Csaki, J. Erlich, J. Terning,
Phys. Rev. D 66 (2002) 064021; \\
%
G. Cacciapaglia, C. Csaki, C. Grojean, J. Terning,
Phys. Rev. D 70 (2004) 075014.
%
\bibitem{CacCsaGroTer05}
G. Cacciapaglia, C. Csaki, C. Grojean, J. Terning,
Phys. Rev. D 71 (2005) 035015.
%
\bibitem{HarMatYam06}
M. Harada, S. Matsuzaki, K. Yamawaki,
Phys. Rev. D 74 (2006) 076004.
%
\end{thebibliography}
\end{document}